\documentclass[10pt]{amsart}
\usepackage{amsmath,amssymb,epsfig,amsthm,amsfonts}
\usepackage{graphicx}

\newtheorem{lemma}[equation]{Lemma}
\newtheorem{proposition}[equation]{Proposition}
\newtheorem{corollary}[equation]{Corollary}

\theoremstyle{definition}

\newtheorem{remark*}{Remark}
\begin{document}
\begin{abstract}In this article we present some results concerning natural dissipative perturbations of 3d Hamiltonian systems. Given a
Hamiltonian system $\dot{x} =PdH$, and a Casimir function $S$,  we construct a symmetric covariant tensor  $g$, so that the modified (so-called ``metriplectic") system $ \dot{x} =PdH+gdS $ satisfies the following conditions: $dH$ is a null vector for $g$, and $dS(gdS)\leq 0$. Along solutions to a dynamical system of this type, the Hamiltonian function $H$ is preserved, while the function $S$ decreases, i.e. $S$ is dissipated by the system. We are motivated by the example of a ``relaxing rigid body" by Morrison \cite{Mor} in which systems of this type were introduced. 
\end{abstract}
\title{Dissipative Perturbations of 3d Hamiltonian Systems}
\author{Daniel Fish}
\address{Department of Mathematics and Statistics, Portland State University, Portland, OR, U.S.}\email{djf@pdx.edu}\date{}
\maketitle

\section*{Introduction}

In his article ``\emph{A Paradigm for Joined Hamiltonian and Dissipative Systems}" (see \cite{Mor}), P.J. Morrison introduced a natural geometric formulation of dynamical systems that exhibit both conservative and nonconservative characteristics. Historically, conservative systems have been modelled geometrically as Hamiltonian systems of the form $\dot{x}=PdH$ where $P$ is a Poisson tensor and $H$ is a smooth function (Hamiltonian function). In a Hamiltonian system, the equality $dH/dt=dHPdH=0$ can be interpreted as conservation of the ``energy" $H$; thus such systems are natural models for conservative dynamics (see \cite{AbMar}). Nonconservative or dissipative systems can also be described geometrically as gradient systems: $\dot{x}=gdS$ where $g$ is a symmetric tensor, $S$ is a smooth function, and $dS/dt=dSgdS$ is typically negative definite. Since the function $S$ is not conserved, it may be interpreted as a form of energy that is dissipated (or as the negative of ``entropy" which is produced by the system) (see \cite{Perko}).

Morrison's formulation combines these two types of systems into a so-called \emph{Metriplectic System}
\[\dot{x}=PdH+gdS,\]
with the additional requirements that $H$ remains a conserved quantity and $S$ continues to be dissipated. These requirements can be met if the following conditions on $H$ and $S$ are satisfied
\[PdS=gdH=0.\]
That is, $S$ is a Casimir function for the Poisson tensor $P$ and $dH$ is a null vector for the symmetric tensor $g$. Morrison applied this formulation to the  equations for the rigid-body with dissipation and the Vlasov-Poisson equations for plasma with collisions.
The formulation of dissipative systems as combined Hamiltonian and gradient systems has also been studied by \cite{BKMR}, \cite{BB}, \cite{KaufTurski},\cite{Kauf}, \cite{Xu} and others in mathematical physics and control theory.

In this article we regard metriplectic systems in $\mathbb{R}^{3}$ as dissipative perturbations of Hamiltonian systems. In the first section we suggest a natural form for the symmetric covariant tensor $g$ that depends only on the differential of the Hamiltonian function $H$, and prove some results about the equilibria of the combined system. In the second section we reproduce Morrison's example as a special case, and present some other interesting applications.

\section{A class of Metriplectic Systems in $\mathbb{R}^{3}$} 

Let $(M,P)$ be a three-dimensional vector space equipped with a Poisson tensor $P$ and standard Euclidean metric $h$. At each point $x$ we identify $T_{x}M$ and $T^{*}_{x}M$ with $\mathbb{R}^{3}$ via the metric $h$: $v^{\sharp}=hv^{\flat}$. When the context is clear, we will denote the dot product on each space, and the pairing between the two spaces (with respect to the metric $h$) by the same symbol, i.e. $u^{\sharp}\cdot v^{\sharp}=u^{\flat}\cdot v^{\flat}=u^{\sharp}\cdot v^{\flat}=u^{i}v_{i}$.

For any function $H$ in $C^{\infty}(M)$, the vector field  $\xi_{P}=PdH$ defines a Hamiltonian system $\dot{x}=\xi_{P}$. Let $S\in C^{\infty}(M)$ be a Casimir function for this system ($PdS=0$). We wish to construct a canonical dissipative perturbation \[ \dot{x} = PdH + gdS, \] of this system, so that the following two conditions hold: $\dot{H} = 0$ and $\dot{S} \leq  0$, where $g$ is a symmetric covariant tensor on $M$. In other words, we want $g$ and $S$ to satisfy
		\begin{equation}\label{typeI} gdH = 0 \quad \textrm{and} \quad dS\cdot gdS  \leq 0.\end{equation}
Let $(x^{1},x^{2},x^{3})$ be local coordinates on $M$, and write $dH=H_{i}dx^{i}$ and $dS=S_{i}dx^{i}$. Assume, for now, that each $H_{i}$ is nonzero. In order for $gdH=0$ to hold, the following relationships between the coefficients of $g$ must be satisfied.
\begin{align}
\notag g^{11} &=-g^{12}(H_{2}/H_{1})-g^{13}(H_{3}/H_{1})\\
\label{diag} g^{22} &=-g^{21}(H_{1}/H_{2})-g^{23}(H_{3}/H_{2})\\		
\notag g^{33} &=-g^{31}(H_{1}/H_{3})-g^{32}(H_{2}/H_{3})
\end{align}
With these diagonal terms, we can calculate $gdS$:
\[gdS=\begin{pmatrix} -(S_{1}/H_{1})(g^{12}H_{2}+g^{13}H_{3})+S_{2}g^{12}+S_{3}g^{13}\\			S_{1}g^{21}-(S_{2}/H_{2})(g^{21}H_{1}+g^{23}H_{3}) + S_{3}g^{23}\\		S_{1}g^{31}+S_{2}g^{32}-(S_{3}/H_{3})(g^{31}H_{1}+g^{32}H_{2})		
		\end{pmatrix} .\]
In each cotangent space $T_{x}^{*}M$ let $\sigma_{i}(x)$ denote the $i^{th}$ component of the cross-product $d_{x}S \times d_{x}H$  and let $\sigma$ be the one-form $\sigma=\sigma_{i}dx^{i}$. Then the vector field $gdS$ can be locally expressed as
\[gdS=\begin{pmatrix} \frac{1}{H_{1}}(g^{13}\sigma_{2}-g^{12}\sigma_{3})\\
		 \frac{1}{H_{2}}(g^{12}\sigma_{3}-g^{23}\sigma_{1})\\
		 \frac{1}{H_{3}}(g^{32}\sigma_{1}-g^{31}\sigma_{2})\end{pmatrix} .\]
Therefore, 
\begin{gather*}
dS\cdot gdS 
=\frac{S_{1}}{H_{1}}(g^{13}\sigma_{2}-g^{12}\sigma_{3}) +\frac{S_{2}}{H_{2}}(g^{12}\sigma_{3}-g^{23}\sigma_{1}) +\frac{S_{3}}{H_{3}}(g^{32}\sigma_{1}-g^{31}\sigma_{2})\\
=\frac{g^{32}\sigma_{1}}{H_{2}H_{3}}(S_{3}H_{2}-S_{2}H_{3})
+\frac{g^{13}\sigma_{2}}{H_{1}H_{3}}(S_{1}H_{3}-S_{3}H_{1})
+\frac{g^{12}\sigma_{3}}{H_{1}H_{2}}(S_{2}H_{1}-S_{1}H_{2})
\\
=-\frac{1}{H_{1}H_{2}H_{3}}(\sigma_{1}^{2}H_{1}g^{32}+\sigma_{2}^{2}H_{2}g^{13}+\sigma_{3}^{2}H_{3}g^{12}).
\end{gather*}
According to the second condition in \eqref{typeI}, we must choose coefficients $g^{ij}$, such that this quantity is non-positive. If we take $g^{ij}=H^{i}H^{j}$ for $i\neq j$ (we have lifted indices of $dH$ via the metric $h$), then we have
\begin{equation} dS\cdot gdS =-(\sigma_{1}^{2}+\sigma_{2}^{2}+\sigma_{3}^{2})=-\left\|\sigma \right\|^{2} \leq 0. \label{sigma}\end{equation}

Substituting $H^{i}H^{j}$ for $g^{ij}$ ($i\neq j$) into \eqref{diag} we find that the diagonal terms of $g$ should have the form $g^{ii}=-\sum_{j\neq i}H^{j}H^{j}$. Thus, we can construct a tensor $g$  that satisfies the required conditions:                         \begin{equation}\label{gI} g^{ij}=H^{i}H^{j}-\delta^{ij}H^{k}H^{k},
\end{equation}
or, invariantly: $g=\nabla H \otimes \nabla H - \textrm{I}\left\| \nabla H\right\|^{2}$, where $\nabla H=dH^{\sharp}$ and $\textrm{I}$ is the unit tensor.
The rank of $g$ is zero only at the points for which $dH=0$, and since $gdH=0$, it is never more than two. In fact, these are the only possibilities.
\begin{lemma} If $d_{x}H\neq 0$ then the tensor $g=\nabla H \otimes \nabla H - \textrm{I}\left\| \nabla H\right\|^{2}$ has rank 2 at the point $x$. \end{lemma}
\begin{proof} Define the following vectors at each point in $M$:
\[v_{1}=(0,H^{3},-H^{2}), \quad v_{2}=(H^{3},0,-H^{1}),\quad v_{3}=(H^{2},-H^{1},0).\]
Observe that at each point in $M$ for which $dH\neq 0$,  the set $\mathcal{J}=\{v_{1},v_{2},v_{3}\}$ spans a subspace of dimension 2. A simple calculation shows that $g(v_{k})^{\flat}=-\left\|dH\right\|^{2}v_{k}$ for each $k$, so the set $\mathcal{J}$ is contained in the image of the homomorphism $\sharp g:T^{*}M\rightarrow TM$. Hence, at points for which $\left\|dH\right\|^{2}\neq 0$, the rank of $g$ is 2. \end{proof}

\begin{figure}[h]
\centering
\epsfxsize=2in \epsfysize=1.5in \epsfclipon
\framebox{\epsfbox[0 500 592 843]{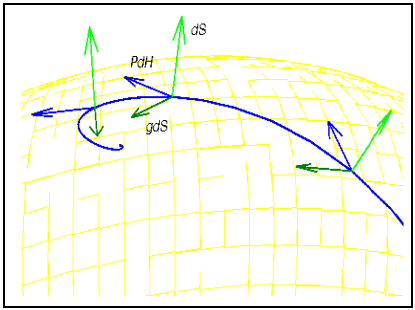}}\caption{Vectors along a solution} \label{path}
\end{figure}

Consider the map $(\sharp g)^{\flat}$ from $T^{*}M\rightarrow T^{*}M$ defined on $v \in T^{*}M$ by lowering an index of $\sharp g(v)$, i.e.  $(\sharp g)^{\flat}(v)=(gv)^{\flat}$. The $k^{th}$ component of $(\sharp g)^{\flat}(dS)$ is 
\begin{align*}(gdS)_{k} &=(H_{k}H^{j}-\delta_{k}^{j}H_{i}H^{i})S_{j}\\
					&= H_{k}H^{j}S_{j}-H_{i}H^{i}S_{k}\\
					&=(dH \cdot dS)H_{k}-(dH \cdot dH)S_{k}\\
					&=[dH\times (dH\times dS)]_{k}.
\end{align*}
Thus, $(gdS)^{\flat}$ can be interpreted geometrically as (a multiple of) the component of $dS$ which is $h$-orthogonal to $dH$, and so $\sharp g$ at the point $x$ can be interpreted as a projection operator from $T^{*}_{x}M$ onto the level set of $H$ that passes through $x$ (see figure \ref{path}). The vector field $PdH$ is a Hamiltonian vector field, and so the vector $Pd_{x}H$ also lies on this level set. Since $gdS$ is constructed to perturb the Hamiltonian system, we would expect $gdS$ and $PdH$ to be independent at most points; in fact we can say more that this.

\begin{lemma}\label{perplemma} At any point in $M$, the vectors $PdH$ and $gdS$ are $h$-orthogonal.\end{lemma}
\begin{proof} Since $P$ is skew-symmetric, and since $S$ is a Casimir for $P$, we have:
\begin{align*}PdH \cdot gdS &= (PdH)^{i} [dH\times (dH\times dS)]_{i}\\
 								&=(PdH)^{i}[(dH\cdot dS)dH - (dH \cdot dH)dS]_{i}\\
 							&=(dH\cdot dS)[PdH \cdot dH]-(dH\cdot dH)[PdH\cdot dS]=0.
\end{align*}
\end{proof}

Using the degenerate covariant metric $g$, we define the metriplectic system

 \begin{equation}\label{sys} \dot{x}=PdH+gdS,\end{equation}
which satisfies $\dot{H}=0$ and $\dot{S}\leq 0$. As for the unperturbed Hamiltonian system, trajectories of \eqref{sys} are contained in level sets of $H$, but they are no longer confined to the symplectic leaves defined by $P$. In fact, by construction, the function $S$ is non-increasing along trajectories: $dS/dt=dS\cdot PdH + dS\cdot gdS=-\left\| \sigma \right\|^{2}$. Notice that this quantity is zero only when $\sigma =0$. That is, the derivative of $S$ along solutions to \eqref{sys} is zero either when $d_{x}S$ is proportional to $d_{x}H$ (i.e., the level sets for $H$ and $S$ are tangent to each other), or when one of $dS$ or $dH$ vanishes. In any case, if $dS/dt$ is zero for all $t$ after some time $T$, then the system has come to ``rest" in the sense that the trajectory is no longer transverse to the symplectic leaves - the dissipation has been turned off. To see this, consider the dissipative part of \eqref{sys}. If $dH=0$, then $g=0$; if $dS=0$, then $gdS=0$; and if $dS$ is parallel to $dH$, then $gdS$ is again zero. In these ``rest states" the system is purely Hamiltonian, and it is to these states that (in most cases) the system will tend.

In order to analyze the behavior of the system as it relaxes to a rest state, as well as its behavior once it has relaxed completely (if it does so), we first make a simple observation regarding the relation between the equilibria of \eqref{sys} and the equilibria of the unperturbed Hamiltonian system. According to lemma \eqref{perplemma}, we know that the vector field $PdH + gdS$ is zero if and only if each component is zero individually. In fact, at most points of $M$, a stronger relation holds.  Recall that a \emph{regular} point for $P$ is a point at which the rank of $P$ is maximal.
\begin{lemma} Define the following vector fields: $\xi_{P}=PdH$ and $\xi=PdH + gdS$. Then, at \emph{regular points} $x\in M$,\, $\xi_{P}(x)=0$ $\leftrightarrow$ $\xi(x)=0$.
\end{lemma}
\begin{proof} If $d_{x}S=0$, then $\xi_{P}(x)=\xi(x)$, so we may assume that $d_{x}S \neq 0$. Similarly, if $d_{x}H=0$, then $g(x)=0$ and again $\xi_{P}(x)=\xi(x)$, so we may further assume that $d_{x}H \neq 0$. 

If $\xi_{P}(x)=0$, then $d_{x}H$ is in the kernel of $P$ which, since $x$ is regular, is spanned by the covector $d_{x}S$. Thus $d_{x}S = \lambda d_{x}H$ and 
				\[\xi(x)=PdH + gdS = \lambda gdH=0.\]
 Conversely, if $\xi(x)=0$, then $0=d_{x}S \cdot \xi(x)=d_{x}S \cdot gd_{x}S=-d_{x}S \times d_{x}H$ (see \eqref{sigma} above). Since both factors of this product are nonzero, they must be proportional: $d_{x}H=\lambda d_{x}S$. But then \[\xi_{P}(x)=PdH=\lambda PdS = 0.\] 
\end{proof}
If $x$ is \emph{not} a regular point for $P$, then $P(x)=0$, and so $\xi_{P}(x)=0$. Such points are clearly equilibria for the Hamiltonian system $\dot{x}=PdH$, but the vector $\xi(x)=gd_{x}S$ may not be zero, so, in general, the conclusion of the above lemma fails to hold in the non-regular case (see \ref{nonreg} below). Although points of degeneracy of $P$ are not typically points of great interest for the Hamiltonian system, they can be relevant to the dynamics of the metriplectic system \eqref{sys}. To distinguish regular and non-regular points of $M$ we define the following set: $\mathcal{R}_{P}=\{x|P(x)\neq 0\}$. Equilibrium points that are in $\mathcal{R}_{P}$ will be called \emph{regular equilibria}. Applying the previous lemma to the two systems $\dot{x}=\xi_{P}$ and $\dot{x}=\xi$ gives us the following
\begin{proposition}\label{p9} The system $\dot{x}=PdH+gdS$ and the unperturbed Hamiltonian system $\dot{x}=PdH$ have the same regular equilibria. \end{proposition}
In other words, perturbing a Hamiltonian system in this way does not alter the regular equilibria of the system. Since the regular equilibria of a Hamiltonian system have a nice geometric description, we can describe fixed points of \eqref{sys} geometrically.
\begin{proposition}\label{Ham} The regular equilibria of the system $\dot{x}=PdH+gdS$ are either critical points of $H$, or are points where the level sets of $S$ and $H$ are tangent to each other.\end{proposition}
\begin{proof} The vector field $\xi_{P}=PdH$ vanishes at a point $x\in \mathcal{R}_{P}$ exactly when either $d_{x}H=0$ or $d_{x}H \in ker(P)$. In the second case, $d_{x}H$ annihilates every vector tangent to the symplectic leaf through $x$. Thus, $d_{x}H$ is normal ($h$-orthogonal) to the symplectic leaf through $x$, which, for $x\in \mathcal{R}_{P}$, coincides with the level set of $S$ through $x$. \end{proof}  
The dissipative system $\dot{x}=gdS$ has its own set of equilibria, but these are not always preserved when the two systems are combined. Points at which the dissipative term $gdS$ vanishes also have a geometric description, related to the extreme values of the function $S$ on the level sets of $H$.
\begin{proposition}\label{p11} The vector fields $\xi=PdH + gdS$ and $\xi_{P}=PdH$ coincide either at critical points of $H$, or at critical points of the function $S$, restricted to a level surface $H=H_{0}$. \end{proposition}
\begin{proof}
The vector field $gdS$ vanishes in one of three cases: $dH=0$, $dS=0$, or $dS, dH \neq 0$ with $dS\in ker(g)$. In the last case, since the rank of $g$ is 2,  the vectors $dS$ and $dH$ must be proportional, i.e. $dS=\lambda dH$. Thus, $\xi(x)=\xi_{P}(x)$ if and only if either $d_{x}H=0$, or $x$ is a critical point for the constrained function $S|_{H_{0}}$.  
\end{proof}
Comparing these results, we obtain the following relation between the equilibria of the component systems $\dot{x}=PdH$ and $\dot{x}=gdS$.

\begin{proposition}If $d_{x}S\neq 0$ at a regular point $x$, then $gd_{x}S=0$ if and only if $Pd_{x}H=0$.\end{proposition} 
\begin{proof} If $Pd_{x}H=0$, then Lemma \ref{perplemma}, together with Prop. \ref{p9}, tells us that $gd_{x}S=0$. On the other hand, from the proof of Prop. \ref{p11} we know that if $gd_{x}S=0$, with $d_{x}S\neq 0$, then $d_{x}H=\lambda d_{x}S$ for some $\lambda \geq 0$, and so $PdH=\lambda PdS=0$.
\end{proof}

\begin{corollary}If $d_{x}S\neq 0$ at a regular point $x$, then $dS/dt = 0$ if and only if $x$ is an equilibrium of \eqref{sys}.\end{corollary}
\begin{proof} If $gd_{x}S=0$, with $d_{x}S\neq 0$, then $dS/dt=dSgdS=0$. The converse follows from Prop. \ref{p9}. \end{proof}

In particular, if $dS\neq 0$ and $P\neq 0$ on a level set $H_{0}$ of $H$, then the only rest states on $H_{0}$ for the system \eqref{sys} are equilibrium points. We also have the following counterpart to Prop. \ref{p9}.

\begin{corollary}A regular point $x$, for which $d_{x}S\neq 0$, is an equilibrium point of the system $\dot{x}=PdH+gdS$ if and only if $x$ is an equilibrium of the gradient system $\dot{x}=gdS$.
\end{corollary}

If $dS=0$ at a regular point $x_{0}$ in $M$, then $PdH$ may or may not vanish (see examples below). The vector field $\xi = PdH + gdS$ reverts to the Hamiltonian one $PdH$ at this point, and $\xi$ is tangent to the level sets $H_{0}$ and $S_{0}$ through $x_{0}$ of both $H$ and $S$. The following result tells us that in this case, $dS=0$ along the entire trajectory $x(t)$ that passes through $x_{0}$, and so $\xi$ remains Hamiltonian along $x(t)$.
\begin{proposition}\label{dS}
Let $x(t)$ be a solution of \eqref{sys} through the point $x_{0}$. If $d_{x_{0}}S=0$, then $d_{x(t)}S=0$ for all time $t$. \end{proposition}
\begin{proof} Let $\mathcal{S}_{0}$ be the symplectic leaf  through $x_{0}$. The one-form $dS$ is constant along $\mathcal{S}_{0}$ since its Lie derivative along any Hamiltonian flow is zero:
\[\mathcal{L}_{PdF}dS=d(\textrm{i}_{PdF}dS)+ \textrm{i}_{PdF}(ddS)=d(dS\cdot PdF)=0,\]
for any smooth function $F$. Since $dS=0$ at the point $x_{0}$, it must remain zero on the entire leaf, and so the vector field $\xi=PdH+gdS$ reduces to $\xi=PdH$ on $\mathcal{S}_{0}$. Thus, the solution $x(t)$ must be contained in the set $\mathcal{S}_{0}$.
\end{proof}
The symplectic leaf $\mathcal{S}_{0}$ through a point $x_{0}$ is a submanifold contained in the level set $S_{0}$ of $S$, with dimension equal to the rank of $P$ at $x_{0}$. When $x_{0}$ is a regular point, the rank of $P$ is 2,  i.e. $\mathcal{S}_{0}=S_{0}$. Thus, when $d_{x_{0}}S=0$, with $x_{0}$ regular, the vector $\xi$ is tangent to $S_{0}$ at every point, and any trajectory of $\eqref{sys}$ that starts on $S_{0}$ must remain there for all time, i.e. the level set $S_{0}$ is invariant under the flow of $\xi$. The system \eqref{sys} restricted to $S_{0}$ is a Hamiltonian system - dissipation is turned off. Moreover, since $dS=0$ along $S_{0}$, the function $S(t)$ is constant along the flow through $x_{0}$. In this case, the system is in a ``rest state"; it is a conservative system with constant (typically maximal or minimal) ``entropy".

\section{Examples and Applications}
In this section we discuss several examples of metriplectic systems in $\mathbb{R}^{3}$ of the type \eqref{sys}. We apply the perturbation method developed in the previous section to some classical Hamiltonian systems.
\subsection{Relaxing Rigid Body (Revisited)}
The example of a ``relaxing rigid body" by Morrison \cite{Mor} was the original motivation for our study of metriplectic systems of this type. We now reproduce this example via an application of the above perturbation method. The Poisson tensor at a point $m=(x,y,z)$ is
\[P=\begin{pmatrix}0&z&-y\\-z&0&x\\y&-x&0\end{pmatrix}.\]
The Hamiltonian and Casimir functions are $H=(1/2)(ax^{2}+by^{2}+cz^{2})$ and $S=(1/2)(x^{2}+y^{2}+z^{2})$. The symmetric tensor $g$ is then
\[g=\begin{pmatrix} -b^{2}y^{2}-c^{2}z^{2}&abxy&acxz\\abxy&-a^{2}x^{2}-c^{2}z^{2}&bcyz\\acxz&bcyz&-a^{2}x^{2}-b^{2}y^{2} \end{pmatrix},\]
which coincides with the operator (up to scalar multiple) defined by Morrison in \cite{Mor}. The vector field $PdH+gdS$ generates a metriplectic system with equations of motion given by
\begin{align*}
		\dot{x} &= (b-c)yz + by(a-b)xy + cz(a-c)xz \\
		\dot{y} &= (c-a)xz + cz(b-c)yz + ax(b-a)xy\\
		\dot{z} &= (a-b)xy + ax(c-a)xz + by(c-b)yx
\end{align*}
Notice that these equations can be expressed as $\dot{m}=PdH + dH\times PdH$ which, since $PdH=dH\times dS$, takes the form
\[\dot{m}=dH\times dS + dH\times (dH\times dS).\]

The only point at which either $dS=0$ or $dH=0$ is the origin, which is a degenerate point for the system. At any other point, the level sets of $H$ and $S$ are ellipsoids and spheres (resp). Along a given level set $H_{0}$ of $H$, the equilibrium points of the system are points where $H_{0}$ is tangent to a sphere, i.e. where $dH$ is parallel to $dS$. 

\begin{figure}[h]
\centering 
\begin{minipage}{2.3in}
\centering 
\epsfxsize=1in \epsfysize=1in \epsfclipon
\framebox{\epsfbox[0 500 592 843]{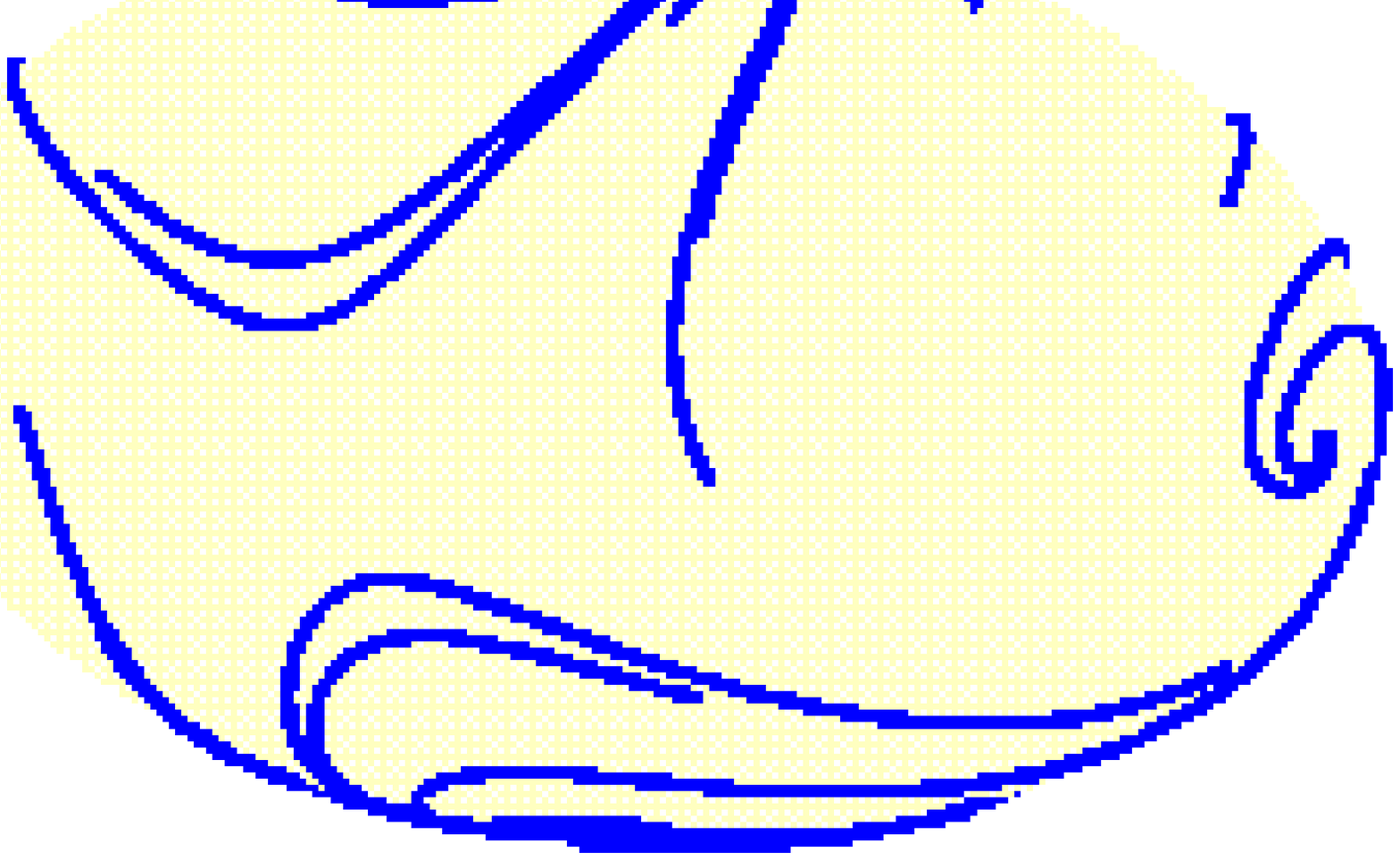}}   \caption{$a,b,c$ distinct}\label{distinct}
\end{minipage}
\hspace{.25in}
\begin{minipage}{2.3in}
\centering
\epsfxsize=1in \epsfysize=1in \epsfclipon
\framebox{\epsfbox[0 500 592 843]{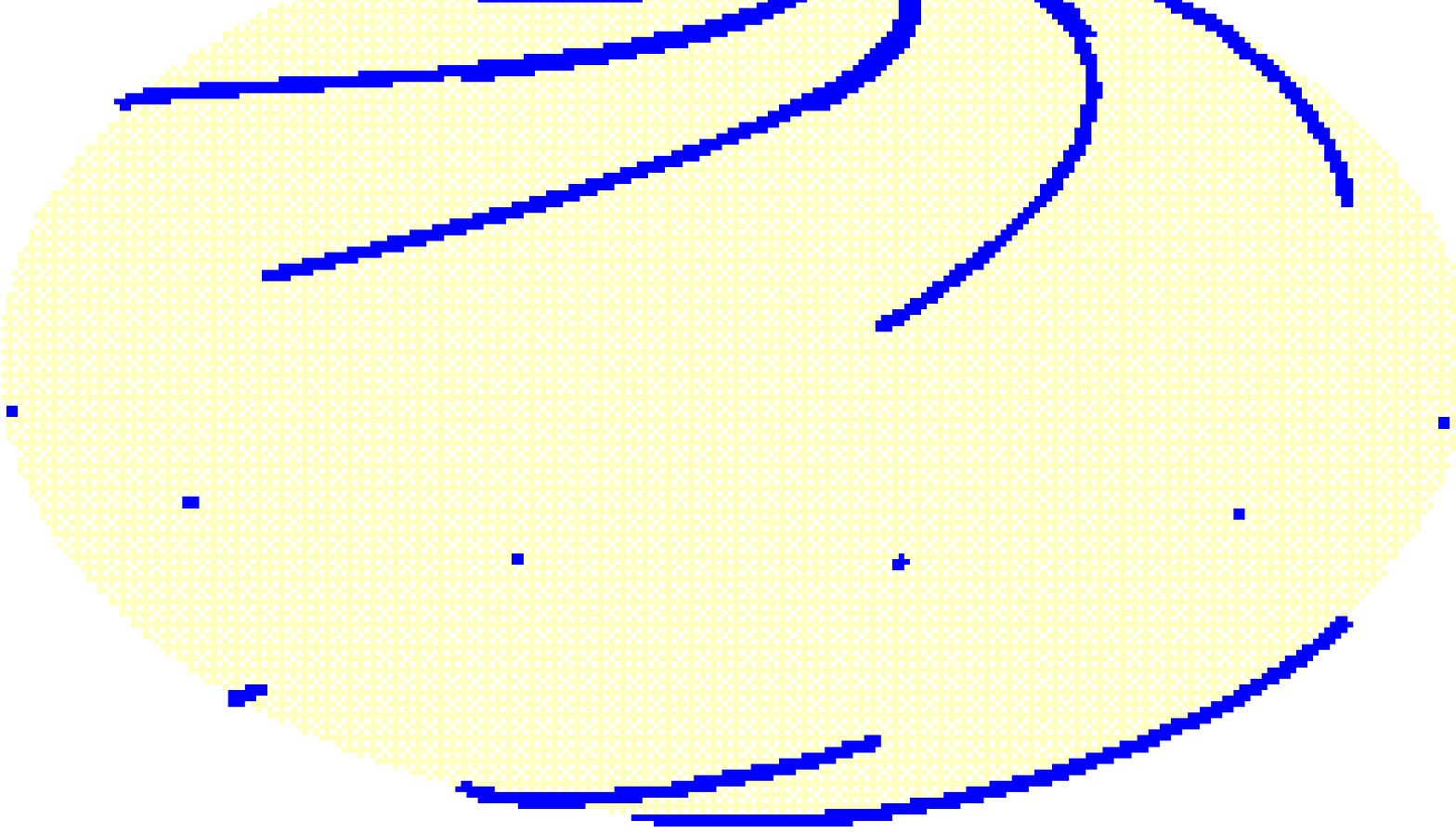}} \caption{$a=b$}\label{twoequal}
\end{minipage} \\
\end{figure}

If $a\neq b\neq c$, then this can only occur at the ``poles" of $H_{0}$ where only one of $x$, $y$, and $z$ is nonzero. At every other point on $H_{0}$, since $dS/dt=-dS\cdot gdS\neq 0$, $S$ must be strictly decreasing . Thus, the pole with the shortest radius on $H_{0}$ is a stable equilibrium, while the other two poles are unstable (see figure \ref{distinct}). If two of $a$, $b$, or $c$ are equal, then two principal radii $r_{i}=r_{j}$ of $H_{0}$ are equal, and the sphere of the same radius is tangent to $H_{0}$ along a circle, every point of which is an equilibrium. The remaining pole is either stable or unstable depending on its length relative to $r_{i}$ (see figure \ref{twoequal}).

\subsection{Dissipative Oscillators}\label{osc}
\begin{figure}[h!]
\centering
\begin{minipage}{2.3in}
\centering
\epsfxsize=1in \epsfysize=1in \epsfclipon
\framebox{\epsfbox[0 300 592 843]{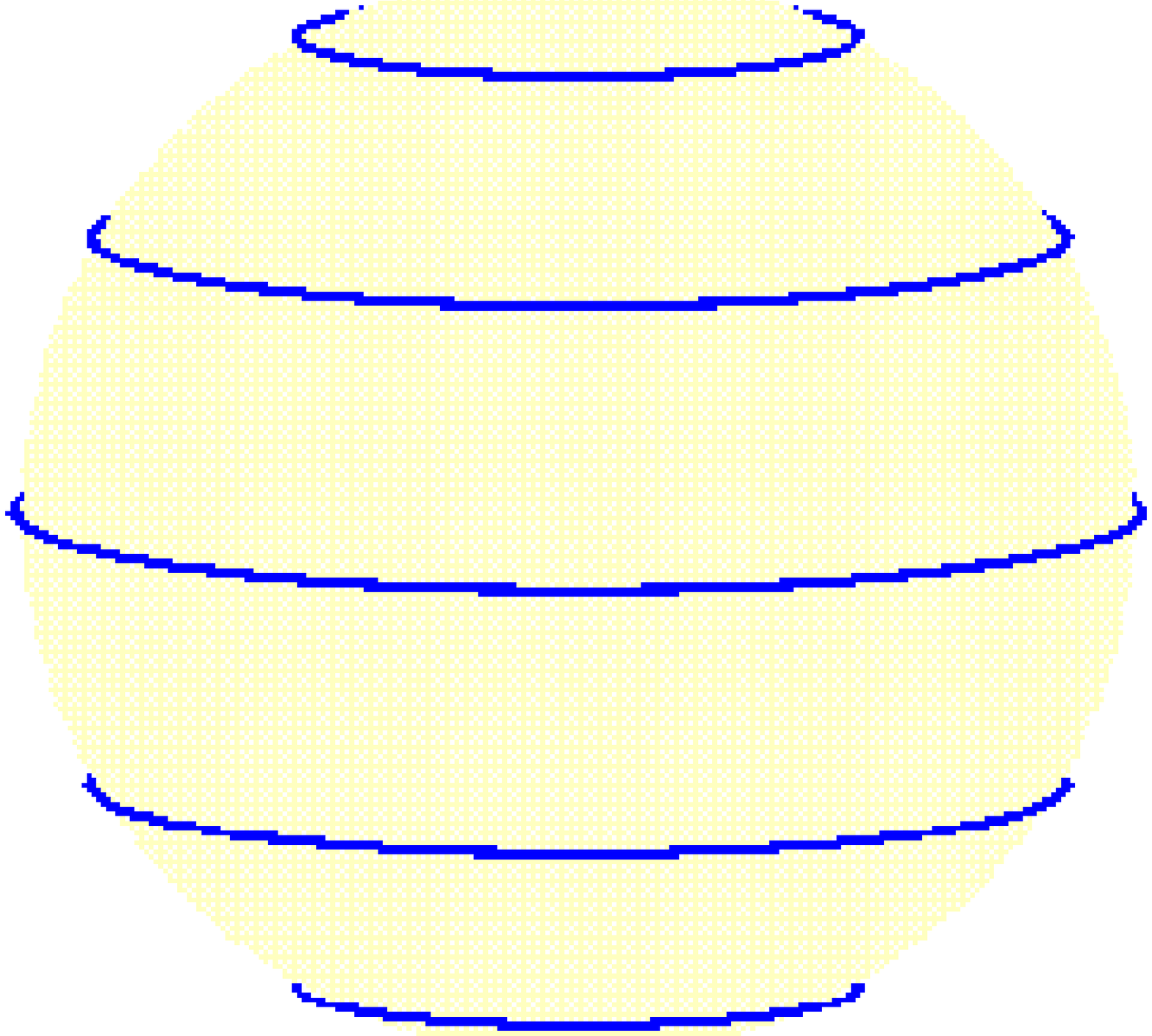}}  \caption{$S=0$}\label{oscfig}
\end{minipage}\hspace{.25in}
\begin{minipage}{2.3in}
\centering
\epsfxsize=1in \epsfysize=1in \epsfclipon
\framebox{\epsfbox[0 260 592 843]{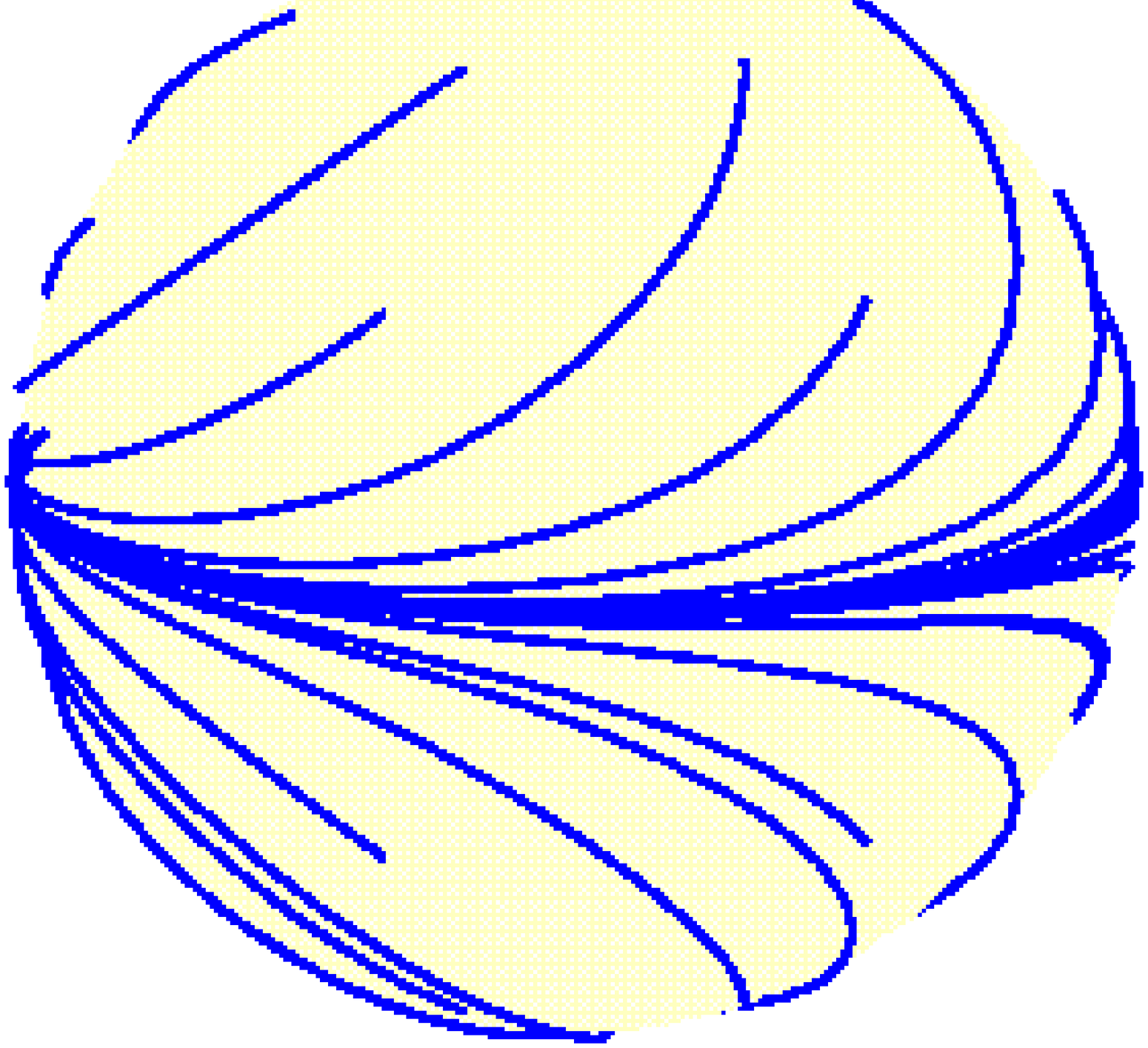}}\caption{$S=z^{2}$} \label{damposc}
\end{minipage} \\
\end{figure}

Here we describe a class of examples in which a one-dimensional oscillator $\ddot{x}=-x$ is perturbed by an external function in the following way. First, write the system as a two-dimensional Hamiltonian system with the standard constant Poisson tensor in $\mathbb{R}^{2}$ and quadratic Hamiltonian function:
\[\begin{matrix} \dot{x} =& y\\ \dot{y} =&-x \end{matrix} \quad \textrm{or} \quad \frac{d}{dt}\begin{pmatrix} x\\y \end{pmatrix} = \begin{pmatrix} 0&1\\-1&0\end{pmatrix} \begin{pmatrix} x\\y\end{pmatrix}
\]
Introduce an external variable $z$ and construct a new 3d Hamiltonian system with Hamiltonian function $H=(1/2)(x^{2}+y^{2}+z^{2})$.
\[ \frac{d}{dt}\begin{pmatrix} x\\y\\z\end{pmatrix} = PdH = \begin{pmatrix} 0&1&0\\-1&0&0\\0&0&0\end{pmatrix} \begin{pmatrix} x\\y\\z \end{pmatrix} = \begin{pmatrix} y\\ -x\\ 0 \end{pmatrix}.\]
A trajectory through $(x_{0},y_{0},z_{0})$ is either the point $(0,0,z_{0})$ on the $z$-axis (equilibrium), or a horizontal circle at height $z_{0}$ that lies on the sphere $H_{0}$ (a level set of $H$) of radius\,  $r=\sqrt{x_{0}^{2}+y_{0}^{2}+z_{0}^{2}}$ \, (see figure \ref{oscfig}). 
The symmetric tensor $g$ that we will use to perturb this system has the form
\begin{equation}\label{gosc} g=\begin{pmatrix} -y^{2}-z^{2}&xy&xz&\\xy&-x^{2}-z^{2}&yz\\xz&yz&-x^{2}-y^{2}\end{pmatrix}.\end{equation}
Now let $S=S(z)$ be any function of $z$, and define the metriplectic system $\dot{m}=PdH+gdS$. In coordinates, 
\begin{align} \dot{x} &= y + xzS'\notag \\
								\dot{y} &= -x + yzS'\label{pend}\\
								\dot{z} &= -(x^{2}+y^{2})S'\notag
\end{align}
where $S'=dS/dz$. The level sets of $S$ are (unions of) horizontal planes, and the differential $dS$ is always parallel to the $z$-axis, except when $S'(z)=0$, in which case $dS=0$. Hence, $dS$ and $dH$ can be parallel only when $dH$ has no horizontal ($x$ or $y$) component, i.e. only at the ``north" and ``south" poles of a level set of $H$.

For a given level set $H_{0}$, the poles $(0,0,\pm z_{0})$, $z_{0}>0$ on the $z$-axis are the only equilibria of \eqref{pend}. The stability of each such point depends on the function $S$. For example, if $S'(z_{0})<0$, then any any solution that begins at a point on $H_{0}$ near $m=(0,0,z_{0})$ will flow toward the pole $m$, and so this point is a stable equilibrium.

If $dS=0$ at some value $z=z_{0}$, then $dS$ will be zero on the whole plane $S_{0}=\{z=z_{0}\}$. Any trajectory $\gamma$ through a point $m$ on $S_{0}$ will remain in $S_{0}$ for all time since the vertical component of its velocity vector will be zero. But $\gamma$ must also remain on the sphere (level set) $H_{0}$ containing $m$, so $\gamma$ lies in the intersection of $S_{0}$ and $H_{0}$. This intersection is either a point (when $m$ is a pole of $H_{0}$), or a circle, in which case $\gamma$ is a cycle on $H_{0}$ at height $z=z_{0}$. Cycles of this type can be either stable or unstable, depending on the function $S$.

For example, if we choose $S(z)=z^{2}$, then the equations of motion become
\begin{align*} \dot{x} &= y + xz^{2}\\
								\dot{y} &= -x + yz^{2}\\
								\dot{z} &= -z(x^{2}+y^{2})
\end{align*}
Since $dS=0$ only at $z=0$, the only periodic solutions are cycles at the equators of spheres $H=k$ of radius $r=\sqrt{2k}$. Clearly $z\rightarrow 0$ along any solution, and $\dot{z}\neq 0$ except at $z=0$. Thus, these equatorial solutions are attractive cycles for the system, and the two poles $(0,0,\pm r)$ of each sphere $H=k$ are unstable equilibria (see figure \ref{damposc}).

\subsection{Points of Degeneracy}\label{nonreg}
As mentioned above, the system $\dot{x}=PdH+gdS$ reduces to the dissipative system $\dot{x}=gdS$ at \emph{non-regular} points, i.e. points where the rank of $P$ is zero. The behavior of the flow at these points varies depending on the choice of Hamiltonian function $H$. Here we present a simple model which displays some of the different possibilities in this situation.

Our model is based on the perturbed oscillator in the previous example, but is altered to allow the rank of $P$ to vanish along the $(x,z)$ plane. Specifically, we define $P$ as
\[P=\begin{pmatrix}0&y&0\\-y&0&0\\0&0&0\end{pmatrix}.\]
For a given function $H$, every point on the plane $\{y=0\}$ is a fixed point for the Hamiltonian system $\dot{m}=PdH$. For any level set $H_{0}=\{H=k\}$ of $H$, the intersection $\mathcal{I}$ of $H_{0}$ and the plane $\{y=0\}$ is invariant under the flow of $\dot{m}=PdH$.

We now perturb this system according to the method described above. The metric $g$ is given by \eqref{gosc}, and the Casimir functions for $P$ are functions that only depend on the third coordinate: $S=S(z)$. The metriplectic vector field in this case has the following form:
\[\xi = PdH+gdS=\begin{pmatrix} yH_{2}+H_{1}H_{3}S'\\ -yH_{1}+H_{2}H_{3}S'\\ -(H_{1}^{2}+H_{2}^{2})S' \end{pmatrix}.\]
Which, when $y=0$, reduces to the vector field $\xi|_{y=0}=gdS$, or
\[\xi|_{y=0}=S'\begin{pmatrix} H_{1}H_{3}\\ H_{2}H_{3}\\ -(H_{1}^{2}+H_{2}^{2}) \end{pmatrix}=S'H_{1}\begin{pmatrix} H_{3}\\ 0\\ -H_{1} \end{pmatrix} + S'H_{2}\begin{pmatrix} 0\\ H_{3}\\ -H_{2} \end{pmatrix}.\]
The points on the plane $\{y=0\}$ are not necessarily fixed by the flow of $\xi$, but the intersection $\mathcal{I}$ will be invariant as long as the vector $gdS|_{y=0}$ is tangent to $\mathcal{I}$, i.e. when the second component of $gdS|_{y=0}$ is zero. From the expression for $\xi$ above, we see that $\mathcal{I}$ is an invariant set exactly when $S'H_{2}H_{3}|_{y=0}=0$. Since $S$ is a function of $z$ only, we have two possibilities: $H_{2}|_{y=0}=0$ or $H_{3}|_{y=0}=0$. In the first case, we have 
\[\xi|_{y=0}=S'H_{1}\begin{pmatrix} H_{3}\\ 0\\ -H_{1}\end{pmatrix}\]
while in the second case:
\[\xi|_{y=0}=S'H_{1}\begin{pmatrix} 0\\ 0\\ -H_{1}\end{pmatrix} + S'H_{2}\begin{pmatrix} 0\\ 0\\ -H_{2}\end{pmatrix}.\]

The following two examples illustrate the behavior of solutions to the equations $\dot{m}=PdH+gdS$. In the first example, the set $\mathcal{I}$ is invariant under the flow of the system, while in the second case $\mathcal{I}$ only remains invariant under the flow of trivial solutions that have initial values on the $(x,y)$ plane.

\begin{figure}[h]
\centering
\begin{minipage}{2.3in}
\centering
\epsfxsize=1in \epsfysize=1in \epsfclipon
\framebox{\epsfbox[0 300 592 843]{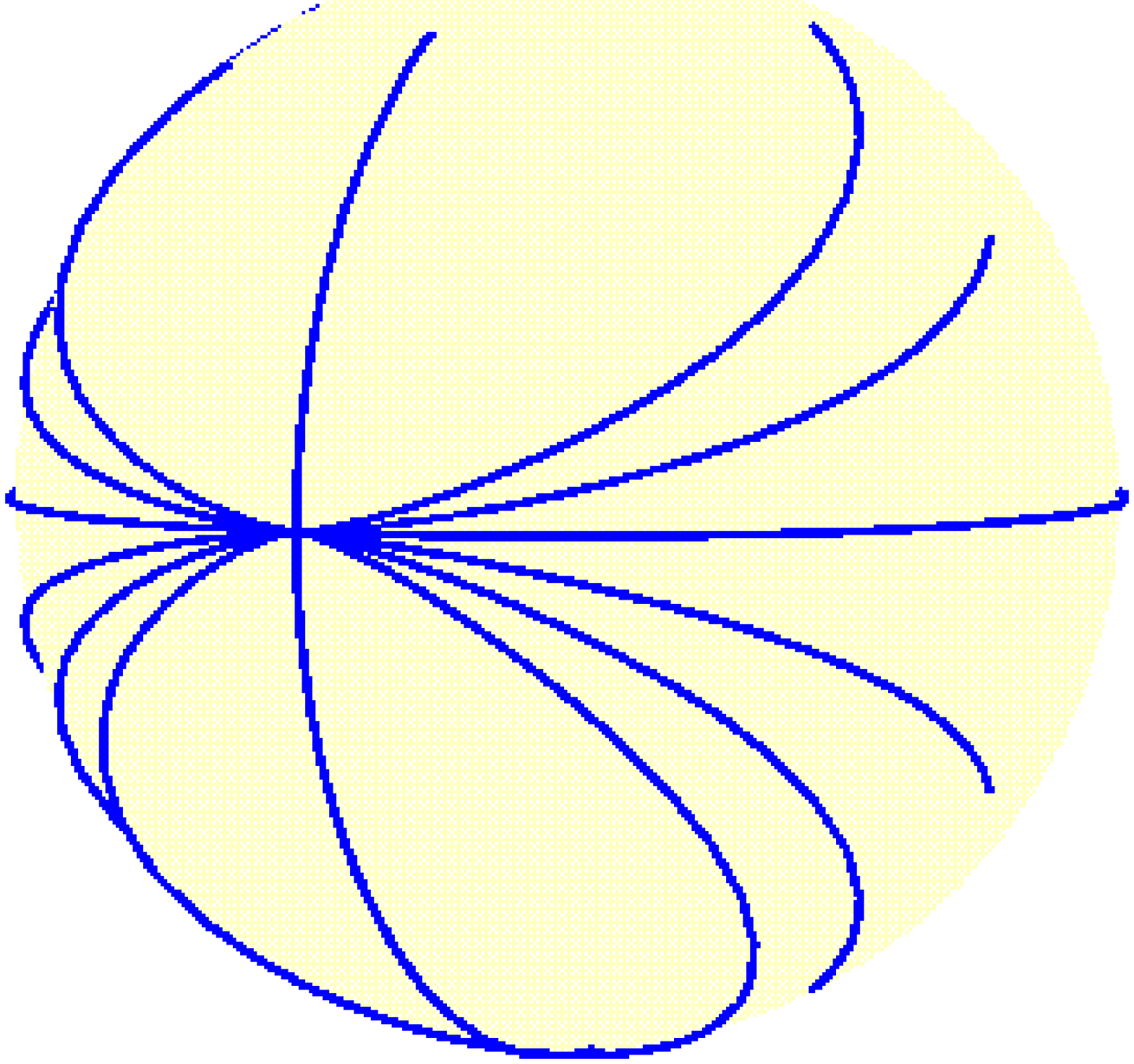}} \caption{$\mathcal{I}$ is invariant} \label{inv}
\end{minipage}\hspace{.25in}
\begin{minipage}{2.3in}
\centering
\epsfxsize=1in \epsfysize=1in \epsfclipon
\framebox{\epsfbox[0 300 592 843 ]{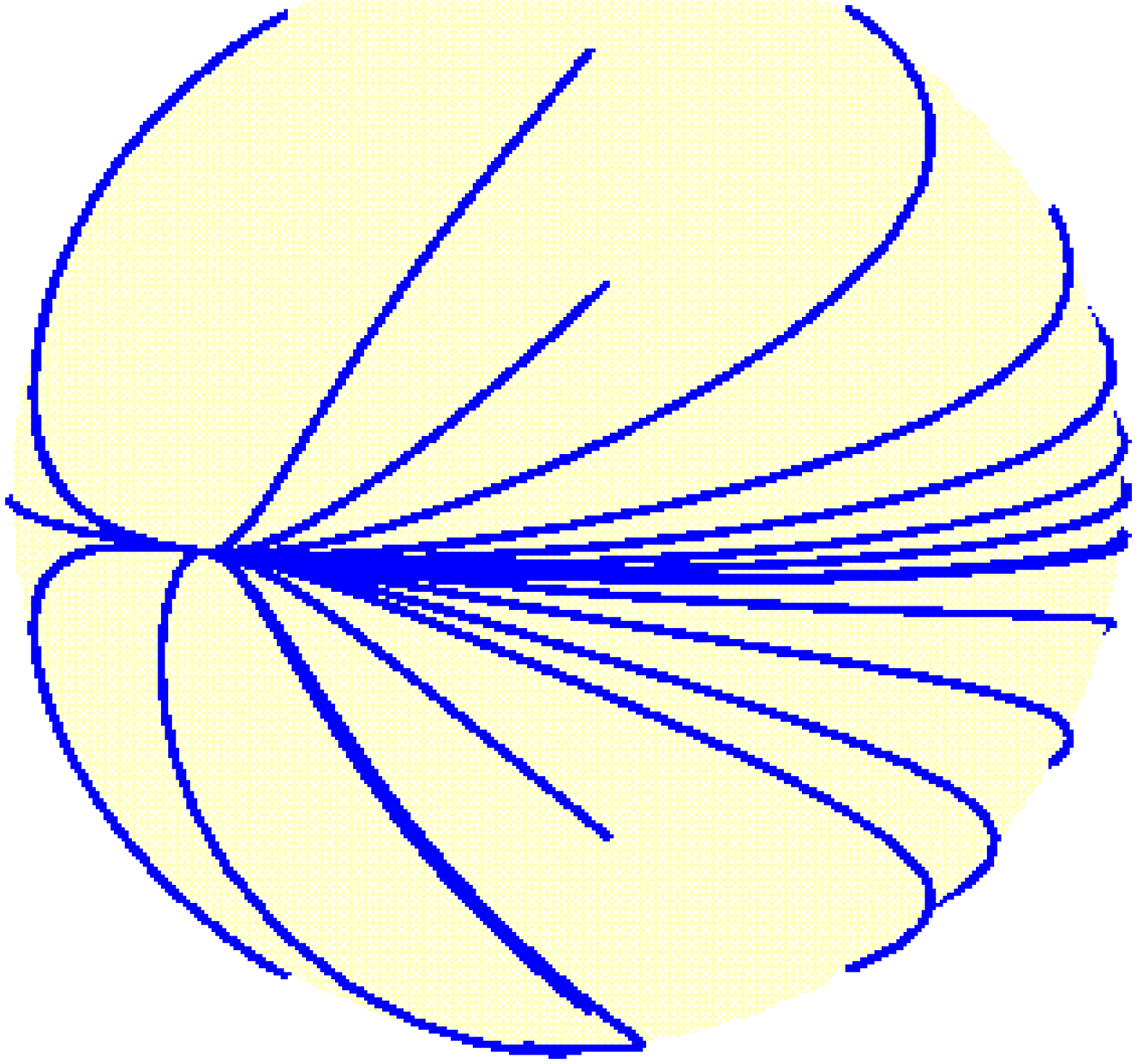}}\caption{$\mathcal{I}$ is not invariant} \label{noninv}
\end{minipage} \\
\end{figure}

\noindent \textbf{Example 1.} Let $H=(x^{2}+y^{2}+z^{2})/2$ and let $S=z^{2}/2$. Then the vector field $\xi$ becomes
\[\xi=PdH+gdS=\begin{pmatrix} y^{2}+xz^{2}\\ -xy + yz^{2}\\ -(x^{2}+y^{2})z\end{pmatrix}.\]
When $y=0$, the tensor $P$ vanishes, and we have
\[\xi|_{y=0}=gdS|_{y=0}=\begin{pmatrix} xz^{2}\\ 0 \\ -x^{2}z \end{pmatrix} = xz\begin{pmatrix} z\\ 0\\ -x\end{pmatrix}.\]
Suppose that $m(t)$ is a solution to $\dot{m}=\xi$ with $m(0)= (x_{0},0,z_{0})$ and let $\mathcal{I}$ be the intersection of the plane $y=0$ with the level set of $H$ that contains $m(0)$. Since the 
$y$-component of $\xi|_{y=0}$ is zero, the point $m(t)$ will be in $\mathcal{I}$ for all time $t$. The set $H_{0}$ is a sphere of radius $r_{0}=\sqrt{x_{0}^{2}+z_{0}^{2}}$, so the set $\mathcal{I}$ is a great circle on this sphere. The points $(0,0,\pm r_{0})$ are unstable equilibria, as in the examples above, but now two new equilibria arise. When the set $\mathcal{I}$ meets the points for which $dS=0$, the vector field $\xi$ vanishes. This occurs at the points $(\pm r_{0},0,0)$ on the equator of $H_{0}$. The point on the positive $x$-axis is a stable fixed point, while the other is unstable (see figure \ref{inv}).

\noindent \textbf{Example 2.} Let $H=(x^{2}+(y-1)^{2}+z^{2})/2$ and let $S=z^{2}/2$. In this case, the vector field $\xi$ is 
\[\xi=PdH+gdS=\begin{pmatrix} y(y-1)+xz^{2}\\ -xy + (y-1)z^{2}\\ -(x^{2}+(y-1)^{2})z\end{pmatrix}.\]
When $y=0$, the tensor $P$ vanishes, and we have
\[\xi|_{y=0}=gdS|_{y=0}=\begin{pmatrix} xz^{2}\\ -z^{2} \\ -(x^{2}+1)z \end{pmatrix} = xz\begin{pmatrix} z\\ 0\\ -x\end{pmatrix}-z\begin{pmatrix}0\\z\\1\end{pmatrix}.\]
The $y$-component of $gdS|_{y=0}$ is zero only if $z=0$. If $m=(x_{0},0,z_{0})$ is a point on the plane $\{y=0\}$, then the flow of $\xi$ through $m$ lies on the level set $H_{0}$ of $H$ containing $m$, but does not remain on the intersection $\mathcal{I}$ unless $z_{0}=0$. The points $(x,0,0)$ where $\mathcal{I}$ meets the $(x,y)$ plane is an equilibrium point, and these are the only points of $\mathcal{I}$ which remain invariant under the flow of $\xi$ (see figure \ref{noninv}).

\section{Summary and Remarks}
In this article we examined metriplectic systems of the type \eqref{sys} from the point of view of perturbations of Hamiltonian systems. We derived a natural form for a symmetric tensor $g$ so that the perturbed system $\dot{x}=PdH+gdS$ dissipates the function $S$ while preserving the energy $H$. We found that a Hamiltonian system $\dot{x}=PdH$ and its metriplectic perturbation have the same \emph{regular} equilibria, which are related to the extreme values of the functions $H$ and $S$. We also found that, for an appropriate choice of $S$, the system \eqref{sys} will tend to a Hamiltonian (possibly equilibrium) rest state in which the dissipative term vanishes. We presented examples of this type of perturbation, including a reproduction of Morrison's example of a `Relaxing Rigid Body'.

\begin{remark*}Although our analysis was restricted to three dimensions, it seems reasonable that certain aspects of our construction should carry over into higher dimensions, including the form of the symmetric tensor $g$. In fact, metriplectic systems have been constructed and discussed in infinite-dimensional settings, examples of which can be found in \cite{Grm2}, \cite{KaufTurski}, and \cite{Mor}.
\end{remark*}

\begin{remark*}The method of perturbation described here can be brought into alignment with the more customary notion of a perturbation by the introduction of a continuous parameter that scales the gradient vector field: $\dot{x}=PdH+ \epsilon gdS$. It would be interesting to study the bifurcations that arise with regard to the stability of equilibria in such systems. For a discussion of such a perturbation involving the Lorenz system, see \cite{Nevir}.
\end{remark*}

\begin{remark*}This article is a product of the author's doctoral dissertation entitled \emph{Metriplectic Systems} which can be found in the archives of the library at Portland State University, or at the link:  web.pdx.edu/$\sim$djf. 
\end{remark*}

\bibliographystyle{plain}
\bibliography{project}
\end{document}